\documentclass[%
 reprint,
superscriptaddress,
 amsmath,amssymb,
 aps,
 prl,
]{revtex4-1}

\usepackage{graphicx}
\usepackage{dcolumn}
\usepackage{bm}

\begin{document}

\preprint{APS/123-QED}

\title{Biological Regulatory Networks are Minimally Frustrated}

\author{Shubham Tripathi}
\affiliation
{
	PhD Program in Systems, Synthetic, and Physical Biology, Rice University, Houston, Texas 77005, USA
}
\affiliation
{
	Center for Theoretical Biological Physics, Rice Univeristy, Houston, Texas 77005, USA
}
\affiliation
{
	Department of Physics, Northeastern Univeristy, Boston, Massachusetts 02115, USA
}

\author{David A. Kessler}
\email
{
	kessler@dave.ph.biu.ac.il
}
\affiliation
{
	Department of Physics, Bar-Ilan Univeristy, Ramat-Gan 52900, Israel
}

\author{Herbert Levine}
\email
{
	h.levine@northeastern.edu
}
\affiliation
{
	Center for Theoretical Biological Physics, Rice Univeristy, Houston, Texas 77005, USA
}
\affiliation
{
	Department of Physics, Northeastern Univeristy, Boston, Massachusetts 02115, USA
}

\date{\today}

\begin{abstract}
Characterization of the differences between biological and random networks can reveal the design principles that enable the robust realization of crucial biological functions including the establishment of different cell types. Previous studies, focusing on identifying topological features that are present in biological networks but not in random networks, have, however, provided few functional insights. We use a Boolean modeling framework and ideas from spin glass literature to identify functional differences between five real biological networks and random networks with similar topological features. We show that minimal frustration is a fundamental property that allows biological networks to robustly establish cell types and regulate cell fate choice, and this property can emerge in complex networks via Darwinian evolution. The study also provides clues regarding how the regulation of cell fate choice can go awry in a disease like cancer and lead to the emergence of aberrant cell types.
\end{abstract}

\maketitle


Biological regulatory networks establish cell type-specific gene expression patterns \cite{hobert_gene_2008} and regulate cell fate-choice in response to various signals. These networks present a contradiction analogous to the famed Levinthal paradox in protein folding \cite{levinthal_how_1969}. Networks as large and complex as those regulating cell fate typically exhibit a huge number of stable states \cite{stein_broken_1989}. Each such stable state or collection of stable states with a reasonably shared pattern of gene expression represents a cell type \cite{huang_regulation_2002, huang_cell_2005}. This relationship, however, predicts a number of cell types much larger than that seen in multicellular organisms. A smaller number of cell types can be attained via the evolutionary fine-tuning of network parameters \cite{kauffman_origins_1993} or by putting cells through a precise sequence of events during development \cite{gilbert_transcriptional_1991-1, *gilbert_transcriptional_1991}. In both scenarios, cell fate will be highly sensitive to intra- and extra-cellular perturbations, an undesirable property.

Features that distinguish biological regulatory networks from random networks may provide a clue regarding how these networks can robustly establish the smaller than expected number of cell types. Biological networks have been shown to often exhibit a scale-free degree distribution \cite{albert_scale-free_2005} which might allow these networks to define topologically stable cell types \cite{aldana_natural_2003}. Regulatory networks in \textit{Escherichia coli} and \textit{Saccharomyces cerevisiae} have been shown to be hierarchically organized \cite{yu_genomic_2006}. Certain network patterns, called motifs, are known to recur far more frequently in biological networks than in random networks \cite{milo_network_2002}, and often mediate cell fate choice \cite{zhou_understanding_2011}. However, these investigations of topological differences between biological and random networks have provided few insights into how the functional characteristics of biological networks differ from those of random networks and allow biological networks to establish cell types. In this Letter, we compare the dynamical behavior of biological networks with that of random networks which have similar topological features and observe some remarkable differences. These could hold the key to elucidating the design principles that allow biological regulatory networks to carry out their biological functions.

\textit{Boolean modeling of biological networks.}--- A Boolean modeling framework \cite{albert_boolean_2014} has proven useful for characterizing the behavior of large networks in cases where the use of ordinary differential equations-based modeling frameworks becomes challenging due to the numerous and hard to estimate kinetic parameters involved. In this framework, the only knowledge required is whether each regulatory relationship between network nodes is activating or inhibitory. The state of a $N$-node network in such a framework may be specified via a sequence $\{s_i\}$ of $N$ binary variables; $s_i=\pm1$. When modeling a biological regulatory network, each network node represents a molecular species such as a transcription factor or micro-RNA. If species $i$ (the molecular species represented by node $i$) is highly expressed, $s_i=+1$, otherwise $s_i=-1$. Regulatory relationships between molecular species are specified by a $N\times N$ matrix $J$ where $J_{ij}=+1$ of species $j$ promotes the expression of species $i$ and $J_{ij}=-1$ if species $j$ inhibits the expression of species $i$. The absence of any regulatory relationship between species $i$ and species $j$ is indicated by $J_{ij}=0$. The discrete-time network dynamics can then be simulated using \cite{font-clos_topography_2018}
\begin{equation}
s_i(t + 1) = \begin{cases}
+1 &\sum\nolimits_{j}J_{ij}s_{j} > 0\\
-1 \quad\text{if}&\sum\nolimits_{j}J_{ij}s_{j} < 0\\
s_i(t) &\sum\nolimits_{j}J_{ij}s_{j} = 0
\end{cases}
\label{eq_one}
\end{equation}
The network state is updated asynchronously, i.e., at each discrete time point, a network node is chosen at random and its state updated using Eq. (\ref{eq_one}). Clearly, a state $\{s_i\}$ is a stable state of the network if $s_i$ is a fixed point of Eq. (\ref{eq_one}) for all $i$.

Note that the dynamical behavior of a network in the above modeling framework is equivalent to the zero-temperature dynamics of an asymmetric spin glass on a graph. Using this equivalence, we characterize an edge $i\rightarrow j$ in state $\{s_i\}$ as frustrated \cite{anderson_concept_1978} if $J_{ij}s_is_j<0$, i.e., if the values of node $i$ and node $j$ in that state do not follow the regulatory relationship between the two nodes. Then, the frustration of a state can be defined as the fraction of network edges that are frustrated in that state. If a network involves regulatory relationships that conflict with one another, all regulatory relationships cannot be satisfied in any state. Hence, such a network will have stable states with non-zero frustration.

\begin{figure}[b]
\includegraphics[width=1.0\linewidth]{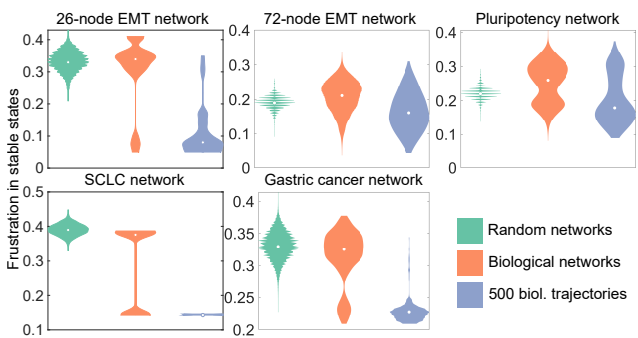}
\caption{\label{fig_1}(Color online) Distribution of frustration for the stable states of biological networks and random networks. The blue (black in print) violin in each panel shows the frustration of the states one ends up in when simulating the dynamics of the biological network starting from 500 random initial conditions. In the case of the 72-node EMT network and in the case of the pluripotency network, 5000 least frustrated stable states of each random network have been shown. The white circle in each violin indicates the median.}
\end{figure}
\textit{Comparison of biological and random networks.}--- We determined the stable states of five biological networks taken from the literature \cite{font-clos_topography_2018, udyavar_novel_2017, jia_testing_2019, chang_systematic_2011, li_endogenous_2015} (see Supplemental Material \footnote{See Supplemental Material file for supplemental figures and detailed methods.}, section II (a)) and compared the frustration of the stable states of each of these networks with the frustration of the stable states of random networks with topological features similar to the biological network (each random network had the same total number of nodes and edges, node in- and out-degree distributions, and the total number of activating and inhibitory relationships between nodes in the network as the corresponding biological network) (Fig. \ref{fig_1}; also see Supplemental Material \cite{Note1}, Fig. S1 and section II (b)-(c)). In the case of each biological network, most stable states had frustration comparable to the frustration of each of the stable states of the corresponding random networks. However, each biological network had a set of stable states with frustration much lower than the frustration of the stable states of random networks. Crucially, biological networks are highly likely to end up in one of these minimally frustrated stable states when their dynamics is simulated starting from random initial conditions (Fig. \ref{fig_1}: blue (black in print) violins in each panel; also see Supplemental Material \cite{Note1}, Fig. S2). Minimally frustrated stable states are thus likely to be biologically significant, with most cells in a population exhibiting gene expression patterns corresponding to these states.

\begin{figure*}
\includegraphics[width=0.75\textwidth]{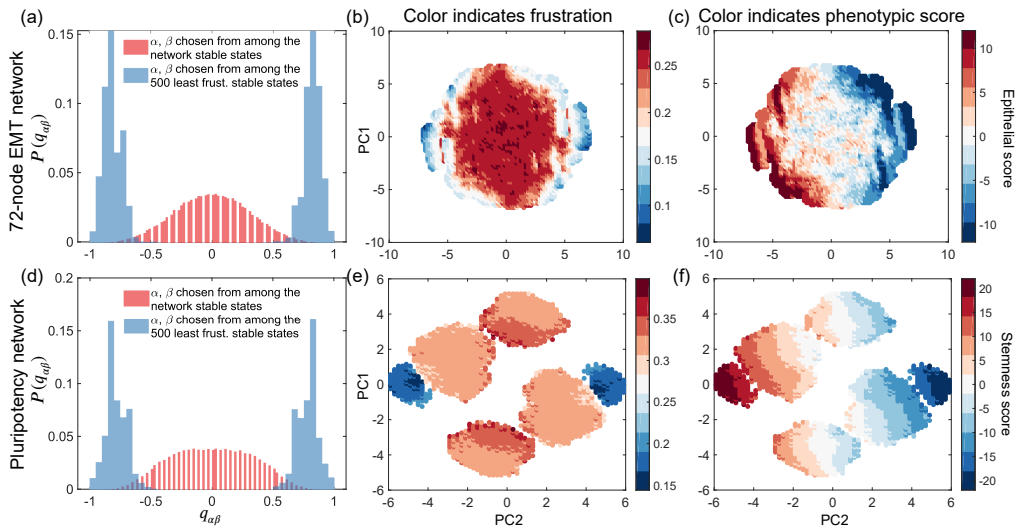}
\caption{\label{fig_2}Minimally frustrated stable states of biological networks define canonical cell types. (a), (d) $P(q_{\alpha\beta})$ is bimodal for the minimally frustrated stable states of both the 72-node EMT network and the pluripotency network. (b), (c), (e), (f) Principal component-representation of the 200000 observed stable states of the 72-node EMT network and of the pluripotency network. In each case, we included 100000 least frustrated and 100000 most frustrated stable states. In (c), a high, positive score indicates an epithelial phenotype while a low, negative score indicates a mesenchymal phenotype. In (f), a high, positive score indicates a stem cell phenotype while a low, negative score indicates a differentiated phenotype. Clusters with extreme values of the phenotypic scores represent canonical cell types. See Supplemental Material \cite{Note1}, section II (d) for details of how the scores were defined.}
\end{figure*}

\textit{Relation between stables states and biological phenotypic states.}--- We next investigated if any structural patterns underlie the organization of stable states of biological networks. The distribution $P(q_{\alpha\beta})$ of the overlap between network states $q_{\alpha\beta}=\sum\nolimits_{i}(s_{i}^\alpha s_{i}^\beta)/N$ was found to be very broad when $\alpha$, $\beta$ pairs are chosen randomly from the set of stable states of biological networks. This indicates a hierarchical organization of stable states \cite{mezard_spin_1986}. However, when the pairs are sampled from among the minimally frustrated stable states, $P(q_{\alpha\beta})$ is bimodal with peaks near $+1$ and $-1$ (Fig. \ref{fig_2} (a) and \ref{fig_2} (d); also see Supplemental Material \cite{Note1}, Fig. S3 (a), S3 (c), and S3 (e)). Since $q_{\alpha\beta}$ is a measure of similarity between the states $\{s_i^\alpha\}$ and $\{s_i^\beta\}$, a collection of states with $q_{\alpha\beta}$ close to $+1$ for all pairs represents a collection of cells with reasonably similar gene expression profiles, to be associated with a distinct cell phenotype. A bimodal $P(q_{\alpha\beta})$ for minimally frustrated stable states of biological networks considered here thus suggests that these states constitute two stable cell phenotypes.

In the case of the network regulating epithelial-mesenchymal transition (EMT) \cite{font-clos_topography_2018}, minimally frustrated stable states represent the two canonical phenotypic states, epithelial and mesenchymal (Fig. \ref{fig_2} (b)-(c); also see Supplemental Material \cite{Note1}, Fig. S3 (b)). In the case of the network regulating pluripotency and differentiation in human embryonic stem cells \cite{chang_systematic_2011}, minimally frustrated stable states define stem and differentiated cell types (Fig. \ref{fig_2} (e)-(f)). In contrast, high frustration stable states of biological networks involve co-presentation of molecular markers corresponding to conflicting biological behaviors. For instance, high frustration stable states of the network regulating EMT involve co-presentation of epithelial and mesenchymal markers (Fig. \ref{fig_2} (b)-(c); also see Supplemental Material \cite{Note1}, Fig. S3 (b)). Similarly, high frustration stable states of the network regulating the neuroendocrine-mesenchymal transition \cite{udyavar_novel_2017} involve co-presentation of neuroendocrine and mesenchymal markers (see Supplemental Material \cite{Note1}, Fig. S3 (d)). These stable states thus represent ambiguous cell fate choices. Such ambiguous phenotypic states have been reported in cancer cells across disease sub-types \cite{udyavar_novel_2017, jolly_implications_2015}, but appear to be suppressed in healthy tissue. To relate more directly to experimental data, we note that deletion of molecular species like GRHL2, OVOL2, and $\Delta$NP63$\alpha$ from the $26$-node EMT network can lower the frustration of network stable states and decrease the fraction of stable states with co-presentation of epithelial and mesenchymal markers (see Supplemental Material \cite{Note1}, Fig. S4). Indeed, loss of expression of such species in cells has recently been shown to inhibit such co-presentation \cite{jolly_stability_2016, watanabe_mammary_2014, dang_np63_2015}.

\begin{figure}[b]
\includegraphics[width=0.75\linewidth]{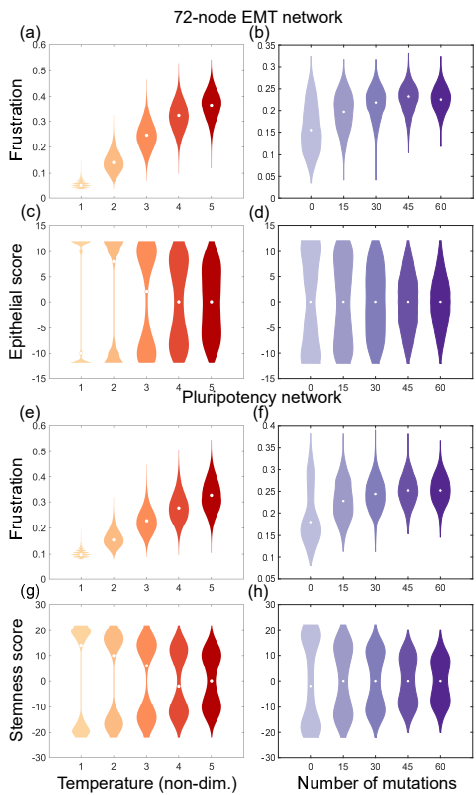}
\caption{\label{fig_3}(Color online) High frustration states are increasingly occupied under noisy dynamics or if the biological network accumulates mutations. (a), (e) Frustration of observed biological network states under noisy node dynamics. The dynamics become more and more noisy as the pseudo-temperature is increased. (b), (f) Frustration of observed states when mutations are introduced into biological networks (without noise in the network dynamics). (c), (d) Epithelial scores of observed states under different levels of noise in the network dynamics (c) and when the network is mutated (d). (g), (h) Stemness scores of observed states under different levels of noise (g) and when the network is mutated (h). The white circle in each violin indicates the median.}
\end{figure}

\textit{Effect of noise in network dynamics.}--- So far, we have neglected stochasticity in our analyses. Noise in gene expression can have significant implications for cellular function \cite{raj_nature_2008}. We defined a pseudo-Hamiltonian $H=-\sum\nolimits_{i,j}J_{ij}s_is_j$ and used the finite-temperature Metropolis Monte Carlo algorithm \cite{metropolis_equation_1953, hastings_monte_1970} to probe network behavior under noisy dynamics (for details of the simulations, see Supplemental Material \cite{Note1}, section II (e)). As node dynamics become increasingly noisy, biological networks become more and more likely to exhibit states with high frustration (Fig. \ref{fig_3} (a) and \ref{fig_3} (e); also see Supplemental Material \cite{Note1}, Fig. S5 (a)-(c), and S6). Functionally, this manifests as more and more cells in a population presenting with ambiguous cell fate choices (Fig. \ref{fig_3} (c) and \ref{fig_3} (g)).

\textit{Effect of network mutations.}--- Another scenario in which cells presenting ambiguous cell fate choices are frequently observed in our modeling framework is if the biological network becomes mutated (Fig. \ref{fig_3} (b) and \ref{fig_3} (f); also see Supplemental Material \cite{Note1}, Fig. S5 (d)-(f), and S7). We observed that the studied biological networks are relatively robust, and it is only after a signifivant number of mutations have accumulated that a significant fraction of cells in the population start exhibiting non-canonical phenotypic states. Since high frustration stable states are far more numerous than minimally frustrated stable states, cell-to-cell variation in network states will be higher when cells exhibit high frustration.

\begin{figure}[b]
\includegraphics[width=1.0\linewidth]{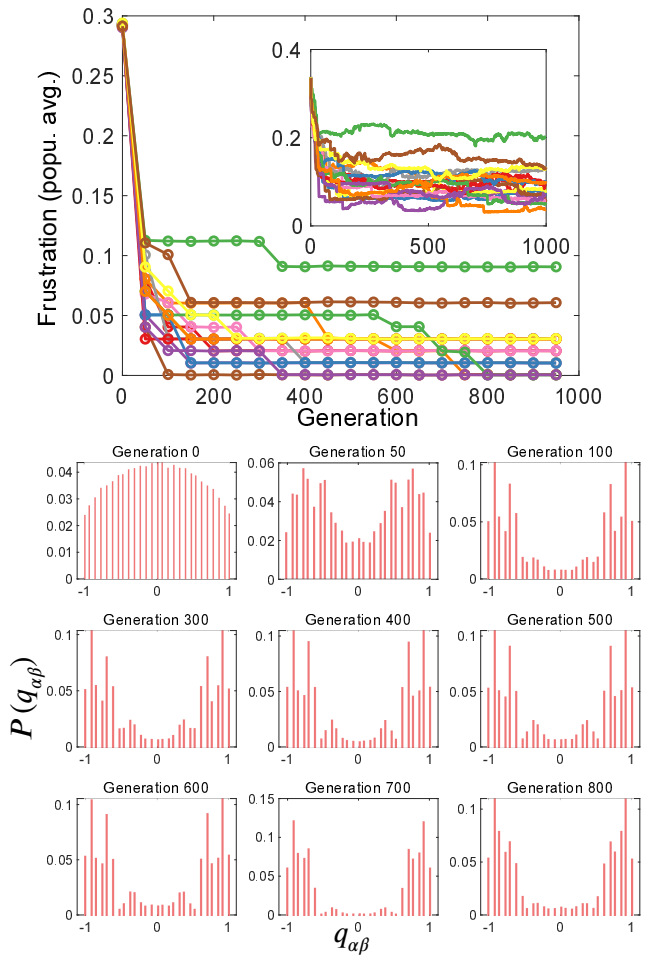}
\caption{\label{fig_4}(Color online) Evolution of biological behavior by a population of random networks under selection for networks with low frustration states. (Top) Frustration of the least frustrated observed state averaged over the networks in a population of 500 networks. Different curves indicate independent simulation runs. (Inset) State frustration averaged over the end states of simulations starting from 50 random initial conditions for each network followed by averaging over the networks in the population. The initial population of 500 random networks was generated from the 26-node EMT network. (Bottom) $P(q_{\alpha\beta})$ at different time points during the evolution simulation, shown for one of the simulation runs.}
\end{figure}
\textit{Emergence of biological characteristics in random networks.}--- Under selection for networks with low frustration states, a population of randomized $26$-node EMT networks can evolve to exhibit the behavior reported herein for the corresponding biological network (Fig. \ref{fig_4}; see Supplemental Material \cite{Note1}, section II (f) for details of the simulation). This includes existence of minimally frustrated stable states that are frequently encountered when starting from random initial conditions (Fig. \ref{fig_4} (Top)) and a bimodal $P(q_{\alpha\beta})$ when $\alpha$, $\beta$ pairs are sampled from among the minimally frustrated stable states (Fig. \ref{fig_4} (Bottom)). That such an evolutionary process is feasible lends crucial support to the hypothesis that the existence of minimally frustrated stable states is a feature that has been acquired by complex biological networks over evolutionary time. Finally, preliminary data suggests that one can relax the assumption of a Boolean modeling framework without changing any of the conclusions (see Supplemental Material \cite{Note1}, section II (g), Fig. S8, S9 (b), and S9 (d)).

\textit{Discussion}--- In the energy landscape description of protein folding \cite{bryngelson_spin_1987, bryngelson_funnels_1995}, existence of minimally frustrated structural conformations distinguishes biological proteins from random heteropolymers. Here, we have shown that the existence of minimally frustrated stable states similarly distinguishes biological regulatory networks from random networks. These minimally frustrated stable states represent canonical cell types and because most random initial conditions dynamically evolve to one of the minimally frustrated stable states, biological networks can robustly establish cell types and regulate cell fate choice between these types. The number of commonly observed cell fates is thus limited to the number of expression patterns in these minimally frustrated stable states. The minimal frustration property distinguishes stable states corresponding to canonical cell types from other possible stable states of the biological network. In contrast, while a random network may have a collection of stable states with an expression pattern similar to that of a canonical cell type, these stable states will in no way be special as compared to the numerous other stable states the random network can exhibit.

Cancer cells exhibit very noisy gene expression which may be driven by the overexpression of certain genes \cite{hinohara_kdm5_2018, domingues_loss_2018, yamamoto_jarid1b_2014}, corrupted epigenetic regulation \cite{pastore_corrupted_2019}, or by metabolic re-programming \cite{carrer_acetyl-coa_2019}. Our results suggest that given the high gene expression noise, cancer cells must frequently exhibit ambiguous cell fate choices. Such behavior has been reported across cancer subtypes, and non-canonical phenotypic states in cancer cells have been shown to be associated with disease aggressiveness. For example, hybrid epithelial-mesenchymal cells, reported across cancer types, have been implicated in the metastatic aggressiveness of solid tumors \cite{jolly_implications_2015}. Also, populations of small cell lung cancer cells treated with anti-cancer drugs have been shown to enrich for hybrid neuroendocrine-mesenchymal cells \cite{udyavar_novel_2017}. Lowering of network frustration upon deletion from the EMT network of factors known to stabilize hybrid epithelial-mesenchymal cells (shown in Supplemental Material \cite{Note1}, Fig. S4) further bolsters the evidence for a connection between non-canonical phenotypic states and high frustration in biological networks. Our model thus provides a new perspective on how noise in the dynamics of regulatory networks in cancer cells can contribute towards the failure of anti-cancer therapies--- noise can facilitate the emergence of cancer cells that exhibit non-canonical phenotypic states. Additionally, accumulation of mutations in biological networks, another characteristic associated with cancer progression, will also promote aberrant cell fate choice. Both noisy gene expression and accumulation of mutations have been shown to be key contributors to intra-tumoral heterogeneity, with significant implications for the failure of anti-cancer therapies \cite{marusyk_intra-tumour_2012}. Estimation of network frustration from cancer cell gene expression data will be a direct test of the role of cell fates associated with high frustration states in disease aggressiveness.

\smallskip
This work was supported by the National Science Foundation grants PHY-1427654 and PHY-1935762. D.A.K. acknowledges support from the United States-Israel Binational Science Foundation grant no. 2015619.

\bibliographystyle{apsrev4-2_modified}
\providecommand{\noopsort}[1]{}\providecommand{\singleletter}[1]{#1}%
%


\end{document}